\documentclass{elsart}

\usepackage{adnd}
\usepackage{longtable}

\usepackage{mathptmx}

\usepackage{amsmath}

\usepackage{amssymb,amstext}
\usepackage[pdftex,
    letterpaper=true,
    hyperindex=true,
    breaklinks=true,
    colorlinks=false,
    citecolor=blue,
    pdftitle={},
    pdfauthor={}]
{hyperref}

\usepackage{graphicx}

\begin{document}

\begin{frontmatter}

\journal{Atomic Data and Nuclear Data Tables}

\copyrightholder{Elsevier Science}

\runtitle{Silver}
\runauthor{Schuh}


\title{Discovery of the Silver Isotopes}


\author{A. Schuh},
\author{A. Fritsch},
\author{J.Q. Ginepro},
\author{M. Heim},
\author{A. Shore},
\and
\author{M.~Thoennessen\corauthref{cor}}\corauth[cor]{Corresponding author.}\ead{thoennessen@nscl.msu.edu}

\address{National Superconducting Cyclotron Laboratory and \\ Department of Physics and Astronomy, Michigan State University, \\East Lansing, MI 48824, USA}

\date{June 15, 2009} 

\begin{abstract}
Thirty-eight silver isotopes have so far been observed; the discovery of these isotopes is discussed.  For each isotope a brief summary of the first refereed publication, including the production and identification method, is presented.

\end{abstract}

\end{frontmatter}





\newpage
\tableofcontents
\listofDtables

\vskip5pc

\section{Introduction}\label{s:intro}
In the ninth paper in the series of the discovery of isotopes, the discovery of the silver isotopes is discussed. Previously, the discoveries of cerium \cite{Gin09}, arsenic \cite{Sho09a}, gold \cite{Sch09a}, tungsten \cite{Fri09}, krypton \cite{Hei09}, einsteinium \cite{Bur09}, iron \cite{Sch09b}, and vanadium \cite{Sho09b} isotopes were discussed.  The purpose of this series is to document and summarize the discovery of the isotopes. Guidelines for assigning credit for discovery are (1) clear identification, either through decay-curves and relationships to other known isotopes, particle or $\gamma$-ray spectra, or unique mass and Z-identification, and (2) publication of the discovery in a refereed journal. The authors and year of the first publication, the laboratory where the isotopes were produced as well as the production and identification methods are discussed. When appropriate, references to conference proceedings, internal reports, and theses are included. When a discovery included a half-life measurement, the measured value is compared to the currently adapted value taken from the NUBASE evaluation \cite{Aud03} which is based on the ENSDF database \cite{ENS08}. In cases where the reported half-life differed significantly from the adapted half-life (up to approximately a factor of two), we searched the subsequent literature for indications that the measurement was erroneous. If that was not the case we credited the authors with the discovery in spite of the inaccurate half-life.

\section{Discovery of $^{93-130}$Ag}
Thirty-eight silver isotopes  from A = $93-130$ have been discovered so far; these include two stable, 15 proton-rich and 21 neutron-rich isotopes. According to the HFB-14 model \cite{Gor07}, $^{155}$Ag should be the last particle-stable neutron-rich nucleus. The proton dripline has been reached and it is estimated that five additional nuclei beyond the proton dripline could live long enough to be observed \cite{Tho04}. Thus, about 30 isotopes have yet to be discovered and approximately 55\% of all possible silver isotopes have been produced and identified so far.

\begin{figure}
	\centering
	\includegraphics[width=12cm]{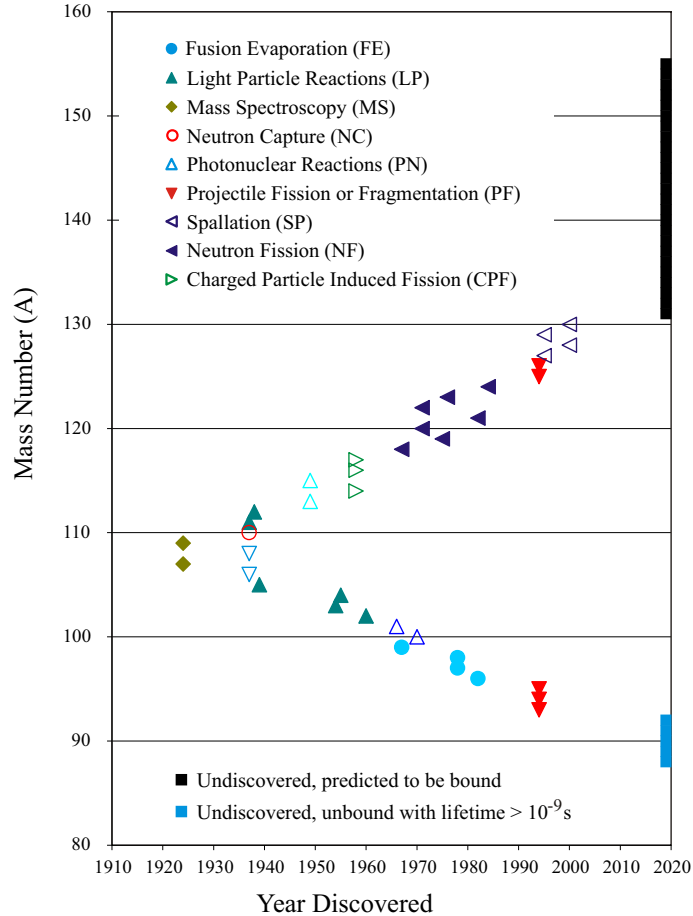}
	\caption{Silver isotopes as a function of time when they were discovered. The different production methods are indicated. The solid black squares on the right hand side of the plot are isotopes predicted to be bound by the HFB-14 model. On the proton-rich side, the light blue squares correspond to unbound isotopes predicted to have lifetimes larger than $\sim 10^{-9}$~s.}
	\label{f:year}
\end{figure}

Figure \ref{f:year} summarizes the year of first discovery for all silver isotopes identified by the method of discovery. The range of isotopes predicted to exist is indicated on the right side of the figure. The radioactive silver isotopes were produced using heavy-ion fusion evaporation (FE), light-particle reactions (LP), neutron-capture reactions (NC), photonuclear reactions (PN), spallation (SP), neutron induced fission (NF), charged-particle induced fission (CPF), and projectile fragmentation or fission (PF). The stable isotopes were identified using mass spectroscopy (MS). Heavy ions are all nuclei with an atomic mass larger than A = 4 \cite{Gru77}. Light particles also include neutrons produced by accelerators. Spallation includes fission induced by high-energy protons. In the following paragraphs, the discovery of each silver isotope is discussed in detail.

\subsection*{$^{93-95}$Ag}\vspace{-0.85cm}

In \textit{Identification of new nuclei near the proton drip line}, Hencheck \textit{et al.} report the discovery of $^{93}$Ag, $^{94}$Ag, and $^{95}$Ag in 1994 \cite{Hen94}. A $^{106}$Cd beam accelerated to 60 MeV/u at the National Superconducting Cyclotron Laboratory (NSCL) at Michigan State University bombarded a natural nickel target. The isotopes $^{93}$Ag, $^{94}$Ag, and $^{95}$Ag were analyzed with the A1900 projectile fragment separator and identified event-by-event with measurements of the magnetic rigidity, time of flight, energy-loss, and total energy. ``A number of new nuclides were identified including $^{88}$Ru, $^{90,91,92,93}$Rh, $^{92,93}$Pd, and $^{94,95}$Ag. A few events corresponding to $^{77}$Y, $^{79}$Zr, $^{81}$Nb, $^{85}$Tc, $^{87}$Ru, $^{91}$Pd, and $^{93}$Ag were also observed.'' Less than three months later, Schmidt \textit{et al.} reported the discovery of $^{94}$Ag and $^{95}$Ag independently \cite{Sch94}.

\subsection*{$^{96}$Ag}\vspace{-0.85cm}

$^{96}$Ag was discovered by Kurcewitz \textit{et al.} in 1982 and reported in their paper \textit{Investigations of Very Neutron-Deficient Isotopes Below $^{100}$Sn in $^{40}$Ca-induced Reactions} \cite{Kur82}. A 4.0 A$\cdot$MeV $^{40}$Ca beam accelerated by the heavy-ion accelerator UNILAC at the Gesellschaft f\"ur Schwerionenforschung (GSI) in Darmstadt, Germany, bombarded a $^{60}$Ni target. $^{96}$Ag was produced in the fusion-evaporation reaction $^{60}$Ni($^{40}$Ca,p3n) and identified by its $\beta$-delayed proton decay: ``The proton activity observed at mass 96 was assigned to $^{96}$Ag from considerations including predicted mass-excess, formation cross-section and the Q-value based preference for odd-N precursors.'' The measured half-life of 5.1(4)~s is consistent with the currently accepted value of 4.45(6)~s.

\subsection*{$^{97,98}$Ag}\vspace{-0.85cm}

$^{97}$Ag and $^{98}$Ag were discovered by Huyse \textit{et al.} in 1978 as described in the paper \textit{The Decay of Neutron Deficient $^{97}$Ag, $^{98}$Ag and $^{99m}$Ag} \cite{Huy78}. A $^{92}$Mo target was irradiated with a 110 MeV $^{14}$N beam from the CYCLONE cyclotron at Louvain-la-Neuve, Belgium. $^{97}$Ag and $^{98}$Ag were identified with Leuven-Isotope-Separator-On-Line (LISOL) and various Ge(LI) $\gamma$- and x-ray detectors. ``Therefore we postulate that the 686.2- and 1294.1-keV $\gamma$ rays originate in the decay of $^{97}$Ag... The presence of [$^{97}$Pd], though not necessarily fed in the $\beta$ decay, suggests J$^{\pi}$ = 6$^{+}$ or 7$^{+}$ for the 44.5-sec $^{98}$Ag that we observe.'' The half-life of $^{97}$Ag was determined to be 21(3)~s which is consistent with the accepted value of 25(3)~s. The half-life for $^{98}$Ag was 44.5(12)~s which is close to the accepted value of 47.5(3)~s. An earlier reported half-life for $^{97}$Ag of 3~m could not be confirmed \cite{Lip69}.

\subsection*{$^{99}$Ag}\vspace{-0.85cm}

\textit{Decay of the isomeric states of $^{102}$Ag} reported the discovery of $^{99}$Ag by Bakhru \textit{et al.} in 1967 \cite{Bak67}. Beams of $^{11}$B from the Yale Heavy Ion Accelerator were incident on natural molybdenum targets and $^{99}$Ag was produced in a fusion-evaporation reaction. The resulting activities were measured with Li-Ge detectors and scintillation counters. ``During these experiments a positive identification of 10$\pm$1 min $^{101}$Ag, 8$\pm$1 min $^{100}$Ag and 3$\pm$0.5 min $^{99}$Ag activities has been made.'' The half-life of 3.0(5)~m for $^{99}$Ag is in reasonable agreement with the currently accepted value of 124(3)~s for an isomeric state. Four months later, an independent measurement reported a half-life of 106(10)~s \cite{Dot67}, however, based on the $\gamma$-ray energies measured coincidences it was later speculated that this measurement corresponded probably to $^{100}$Ag \cite{Hna70}.

\subsection*{$^{100}$Ag}\vspace{-0.85cm}

Hnatowich \textit{et al.} correctly identified $^{100}$Ag for the first time in 1970 as reported in \textit{The decay of Cadmium isotopes of mass 100, 101, and 102 to isomers in silver} \cite{Hna70}. A high purity molten tin target was irradiated with 600 MeV protons from the CERN 600 MeV Synchro-cyclotron and $^{100}$Cd was produced in the Sn(p,3pxn) spallation reaction. $^{100}$Ag was then observed in the ISOLDE isotope separator facility. ``The $^{100}$Cd does not decay appreciably to the previously known 8 min $^{100}$Ag but, instead, to an isomer of half-life 2.3$\pm$0.1 min.'' This value for the half-life agrees with the presently accepted value of 2.01(9)~m. The incorrect half-life of about 8~m had previously been reported by several authors \cite{Bak67,Dot67,But66}.

\subsection*{$^{101}$Ag}\vspace{-0.85cm}

In the paper \textit{Neutron-Deficient Silver and Cadmium Isotopes}, Butement and Mirza described the observation of $^{101}$Ag in 1966 \cite{But66}. 340 MeV protons from the University of Liverpool Synchocyclotron bombarded a silver wool target. ``The half life of $^{101}$Ag was determined by preparing by spallation a pure silver activity 25 min after the end of irradiation and milking off palladium at regular intervals which varied from 7 to 20 min in different experiments.'' A half-life of 14~m was determined which is consistent with the currently adapted value of 11.1(3)~m and corresponds to an isomer of $^{101}$Ag. About four months later, Panontin and Caretto reported a half-life of 11.2(1)~m \cite{Pan66}. They were apparently not aware of the data by Butement and Mirza, however they refer to a conference contribution by Charoenkwan \textit{et al.} as the first observation of $^{101}$Ag \cite{Cha65}.

\subsection*{$^{102}$Ag}\vspace{-0.85cm}

In their 1960 paper \textit{Spins and Decay Modes of Certain Neutron-Deficient Silver Isotopes}, Ames \textit{et al.} identify $^{102}$Ag correctly for the first time \cite{Ame60}. $^{102}$Ag was produced by bombarding a $^{102}$Pd target with 18-MeV protons from the Princeton University cyclotron. The activities were measured with a NaI crystal x-ray counter. ``The present work appears to provide the first direct evidence for the existence of a 15-min activity in Ag$^{102}$.'' This half-life (15(2)~m) agrees with the presently accepted value of 12.9(3)~m. The observation of $^{102}$Ag had actually been reported more than 21 years earlier. Although the abstract of the paper by Enns indicates a correct half-life measurement ``...and three new periods of 16.3 min (+), 73 min. (+) and 45 days (K capture). The latter are assigned tentatively to Ag$^{102}$, Ag$^{104}$, and Ag$^{105}$, respectively,'' the paper itself clearly assigns the 73~min half-life incorrectly to $^{102}$Ag \cite{Enn39}. It is interesting to note that this assignment was reversed in the 1958 Table of Isotopes \cite{Str58}.

\subsection*{$^{103}$Ag}\vspace{-0.85cm}

Haldar and Wiig reported the discovery of $^{103}$Ag in their 1954 article \textit{New Neutron-Deficient Isotope of Iron}   \cite{Hal54}. Silver was bombarded with high-energy protons from the University of Rochester 130-inch synchrocyclotron. Following chemical separation, the radioactive decay was measured with an x-ray proportional counter. ``... in view of the relatively short half-life of Ag$^{104}$ and of the energy available in the transition from the ground state of Ag$^{102}$ to that of Pd$^{102}$, Ag$^{102}$ should have a short half-life. This suggested that our observed 1.1-hour Ag activity was due to Ag$^{103}$, a conclusion which was confirmed by extraction of the known 17-day Pd$^{103}$ daughter.''
This half-life is consistent with the adapted value of 65.7(7)~m. A half-life of 1.1~h had been previously been observed by Bendel \textit{et al.}, however, they assigned the decay to either $^{102}$Ag or $^{104}$Ag assuming it corresponded to the 73~m half-life of Enns \cite{Enn39}.

\subsection*{$^{104}$Ag}\vspace{-0.85cm}

In \textit{The New Isotopes Cd$^{104}$ and Ag$^{104}$}, Johnson reported the observation of $^{104}$Ag in 1955 \cite{Joh55}. Protons were accelerated to 50 MeV by the McGill University 100 MeV synchrocyclotron and bombarded metallic silver. $^{104}$Ag was studied following the decay of $^{104}$Cd which was produced in the reaction $^{107}$Ag(p,4n) with a 180-degree spectrograph, a lens spectrometer and a scintillation spectrometer. ``Since conversion lines had already been found of half-life $\sim$59 min., and these showed no growth, it was evident that the 27 min. activity was a daughter product of the 59 min. activity (Cd$^{104}$) and should therefore be assigned to Ag$^{104}$.'' The 27~m half-life is consistent with the adapted value of a 33.5(2)~m isomer. Earlier measurements of a half-life of about 70~m which corresponds to the ground state of $^{104}$Ag were inconclusive and were not uniquely assigned to $^{104}$Ag (See $^{102}$Ag) \cite{Enn39,Lin50}.

\subsection*{$^{105}$Ag}\vspace{-0.85cm}

The first observation of $^{105}$Ag was reported by Enns in 1939 as described in \textit{Radioactivities Produced by Proton Bombardment of Palladium} \cite{Enn39}. Palladium targets were bombarded with fast protons at the University of Rochester. The decay curves were measured for x-rays, $\gamma$-rays, and conversion electrons. ``Considering the possible products of p-n reactions, Ag$^{105}$ was the unassigned isotope of odd mass number closest to the stable Ag isotopes. Hence the longest of the periods was assigned to it.'' This measured half-life of 45~d is consistent with the presently accepted value of 41.29(7)~d. The observation of a 7.5~d half-life reported in a conference proceeding \cite{Kra37} had been incorrectly assigned to $^{105}$Ag in the 1937 review of Nuclear Physics \cite{Liv37}.

\subsection*{$^{106}$Ag}\vspace{-0.85cm}

Bothe and Gentner first identified $^{106}$Ag in their 1937 paper \textit{Herstellung neuer Isotope durch Kernphotoeffekt} \cite{Bot37}. $^{106}$Ag was produced in the reaction $^{107}$Ag($\gamma$,n): ``Silber zeigte eine neue Halbwerts\-zeit von 24 min. Von den beiden bekannten, durch Neutronenanlagerung entstehenden Halbwertzeiten wurde au\ss erdem die von 2.3 min erhalten, nicht aber die von 22 sec. Hiernach ist folgende Zuordnung anzunehmen: Ag$^{106}$ = 24 min; Ag$^{108}$ = 2.3 min; Ag$^{110}$ = 22 sec.'' (Silver showed a new half-life of 24 min. In addition, of the two known half-lives produced by neutron addition, the 2.3 min half-life was observed, however, not the 22 sec half-life. Therefore, the following assignment can be assumed: Ag$^{106}$ = 24 min; Ag$^{108}$ = 2.3 min; Ag$^{110}$ = 22 sec.). The half-life for $^{106}$Ag agrees with the currently accepted value of 23.96(4)~m. This assignment was confirmed several times in the same year \cite{Poo37,Cha37,Rot37,Kra37}.

\subsection*{$^{107}$Ag}\vspace{-0.85cm}

Aston identified the stable isotope $^{107}$Ag in 1924 in \textit{The Mass Spectra of Chemical Elements - Part V} \cite{Ast24}. $^{107}$Ag was identified using silver cloride and lithium chloride anodes for the mass spectrometer in Cambridge, England. ``Silver has two isotopes, whose masslines when measured against that of iodine have integral values 107 and 109.''

\subsection*{$^{108}$Ag}\vspace{-0.85cm}

Bothe and Gentner first identified $^{108}$Ag in their 1937 paper \textit{Herstellung neuer Isotope durch Kernphotoeffekt} \cite{Bot37}. $^{108}$Ag was produced in the reaction $^{109}$Ag($\gamma$,n): ``Silber zeigte eine neue Halbwerts\-zeit von 24 min. Von den beiden bekannten, durch Neutronenanlagerung entstehenden Halbwertzeiten wurde au\ss erdem die von 2.3 min erhalten, nicht aber die von 22 sec. Hiernach ist folgende Zuordnung anzunehmen: Ag$^{106}$ = 24 min; Ag$^{108}$ = 2.3 min; Ag$^{110}$ = 22 sec.'' (Silver showed a new half-life of 24 min. In addition, of the two known half-lives produced by neutron addition, the 2.3 min half-life was observed, however, not the 22 sec half-life. Therefore, the following assignment can be assumed: Ag$^{106}$ = 24 min; Ag$^{108}$ = 2.3 min; Ag$^{110}$ = 22 sec.). The half-life for $^{108}$Ag agrees with the currently accepted value of 2.37(1)~m. Half-lives of 2~m \cite{Fer34} and 2.3~m \cite{Ama35} had been previously reported for silver, however, no mass assignment was made. The assignment was also confirmed two more times in the same year \cite{Poo37,Kra37}.

\subsection*{$^{109}$Ag}\vspace{-0.85cm}

Aston identified the stable isotope $^{109}$Ag in 1924 in \textit{The Mass Spectra of Chemical Elements - Part V} \cite{Ast24}. $^{109}$Ag was identified using silver cloride and lithium chloride anodes for the mass spectrometer in Cambridge, England. ``Silver has two isotopes, whose masslines when measured against that of iodine have integral values 107 and 109.''

\subsection*{$^{110}$Ag}\vspace{-0.85cm}

Bothe and Gentner first identified $^{110}$Ag in their 1937 paper \textit{Herstellung neuer Isotope durch Kernphotoeffekt} \cite{Bot37}. They made the assignment based on the non-observation of $^{110}$Ag in photonuclear reactions on $^{107}$Ag and $^{109}$Ag: ``Silber zeigte eine neue Halbwertszeit von 24 min. Von den beiden bekannten, durch Neutronenanlagerung entstehenden Halbwertzeiten wurde au\ss erdem die von 2.3 min erhalten, nicht aber die von 22 sec. Hiernach ist folgende Zuordnung anzunehmen: Ag$^{106}$ = 24 min; Ag$^{108}$ = 2.3 min; Ag$^{110}$ = 22 sec.'' (Silver showed a new half-life of 24 min. In addition, of the two known half-lives produced by neutron addition, the 2.3 min half-life was observed, however, not the 22 sec half-life. Therefore, the following assignment can be assumed: Ag$^{106}$ = 24 min; Ag$^{108}$ = 2.3 min; Ag$^{110}$ = 22 sec.). The first measurement of 20~s \cite{Fer34} and 22~s \cite{Ama35} half-lives for silver were made by neutron irradiations, however, no mass assignments were made.

\subsection*{$^{111}$Ag}\vspace{-0.85cm}

In \textit{Radioactive Isotopes of Silver and Palladium from Palladium}, Kraus and Cork reported the discovery of $^{111}$Ag in 1937 \cite{Kra37}.  $^{111}$Ag was produced by bombarding palladium with 6.3 MeV deuterons from the University of Michigan cyclotron. Decay curves were measured with Lauritsen quartz fiber electroscopes and a Wulf string electrometer equipped with an ionization chamber following chemical separation. ``If one of the observed periods is due to the isotope of mass 111 then by beta-decay it should produce a radioactive silver since there is no stable silver of mass 111... it appears to be quite certain that this silver activity (180-hr. half-life) must be built up from the 17-min. and not the 13-hr. palladium.'' The half-life of 180~h (7.5~d) is consistent with the accepted value of 7.45(1)~d.

\subsection*{$^{112}$Ag}\vspace{-0.85cm}

The radioactive isotope $^{112}$Ag was first produced by Pool in 1938 and reported in the article \textit{Radioactivity in Silver Produced by Fast Neutrons} \cite{Poo38}.  Metallic cadmium and indium targets were bombarded with fast neutrons from the Li+H$^{2}$ reaction at the University of Michigan cyclotron. Following chemical separation, the activity was measured with a Wulf string electrometer equipped with an ionization chamber. ``Since this period can be obtained only from indium and cadmium, it seems most probable that silver, $^{112}$Ag, is the carrier of the activity and the reaction equations are as follows: $_{49}$In$^{115}$ + $_0$n$^1 \rightarrow$ $_{147}$Ag$^{112}$ + $_{2}\alpha^4$, and $_{148}$Cd$^{112}$ + $_0$n$^1 \rightarrow$ $_{47}$Ag$^{112}$ + $_1$p$^1$.''  The observed half-life of 3.2(2)~h agrees with the accepted value of 3.130(9)~h.

\subsection*{$^{113}$Ag}\vspace{-0.85cm}

In the 1949 paper \textit{Radioactive Isotopes of Silver Produced by Photo-Disintegration of Cadmium}, Duffield and Knight reported the discovery of $^{113}$Ag \cite{Duf49}. Cadmium oxide enriched with $^{114}$Cd was bombarded with 21~MeV betatron x-rays at the University of Illinois. $^{113}$Ag was produced in the ($\gamma$,p) reaction and identified following chemical separation. ``The silver from the Cd 114 decayed with a half-life of 5.3 hr. over five half-lives, thus establishing that it was Ag 113 made by Cd 114 ($\gamma$,p).'' The extracted half-life agrees with the presently accepted value of 5.37(5)~h. The 1948 Table of Isotopes \cite{Sea48} made the assignment of the 5.3~h half-life based on an unpublished report of the Plutonium Project \cite{Tur48}.

\subsection*{$^{114}$Ag}\vspace{-0.85cm}

Alexander \textit{et al.} discovered in 1958 $^{114}$Ag as reported in \textit{Short-Lived Isotopes of Pd and Ag of Masses 113-117} \cite{Ale58}. Uranium was bombarded with 15 MeV deuterons at Princeton University and the isotopes were produced in the subsequent fission of uranium. $^{114}$Ag was identified following chemical separation by measuring $\beta$-particles and $\gamma$-rays. $^{114}$Ag was identified by a known $^{114}$Cd $\gamma$-ray: ``A level between 0.55 and 0.56 Mev has been found in Cd$^{114}$ by several investigators using Coulomb excitation of Cd$^{114}$Cd and neutron capture of Cd$^{113}$, and it has been found in the decay by K capture of 50-day In$^{114}$. The similarity of the energy levels suggest the mass number 114 for the 2.4-min Pd, 5-sec Ag chain.'' This half-life agrees with the presently accepted value of 4.6(1)~s. The previous observation of a 2~m activity \cite{Duf49} could not be confirmed.

\subsection*{$^{115}$Ag}\vspace{-0.85cm}

In the 1949 paper \textit{Radioactive Isotopes of Silver Produced by Photo-Disintegration of Cadmium}, Duffield and Knight reported the discovery of $^{115}$Ag \cite{Duf49}. Cadmium oxide enriched with $^{116}$Cd was bombarded with 21~MeV betatron x-rays at the University of Illinois. $^{115}$Ag was produced in the ($\gamma$,p) reaction and identified following chemical separation: ``...the 20-min. silver activity was found to be Ag 115 made by Cd 116 ($\gamma$,p).'' The extracted half-life agrees with the presently accepted value of 20.0(5)~m. A previously observed 20~m activity \cite{See46} had been tentatively assigned incorrectly to $^{114}$Ag \cite{See47}.

\subsection*{$^{116,117}$Ag}\vspace{-0.85cm}

Alexander \textit{et al.} discovered in 1958 $^{116}$Ag and $^{117}$Ag as reported in \textit{Short-Lived Isotopes of Pd and Ag of Masses 113-117} \cite{Ale58}. Uranium was bombarded with 15 MeV deuterons at Princeton University and the isotopes were produced in the subsequent fission of uranium. $^{116}$Ag and $^{117}$Ag were identified following chemical separation by measuring $\beta$-particles and $\gamma$-rays. $^{116}$Ag was identified by a known $^{116}$Cd $\gamma$-ray: ``Coulomb excitation of Cd$^{116}$ reveals the presence of a 0.508-Mev level in this nuclide which is, within the experimental error, identical to the $\gamma$ line of 0.515 Mev observed for the 2.5-min Ag. Because of the similarity of these energy levels, it is proposed to assign the 2.5~min Ag to the mass number 116.'' The identification of $^{117}$Ag was determined from the relationship to cadmium decay curves: ``The Cd decay curves of the successive extracts were analysed into components of chains of masses 115 and 117. The data correspond to a half-period of 1.1 min for Ag$^{117}$.'' These half lives are consistent with the presently accepted values of 2.68(10)~m and 73.6(14)~s for  $^{116}$Ag and $^{117}$Ag, respectively. The approximately 3~m half-life had been previously observed, however, no definite mass assignment was made \cite{See46,See47}.

\subsection*{$^{118}$Ag}\vspace{-0.85cm}

In the paper \textit{Identification of 5.3-sec $^{118}$Ag as a Product of $^{238}$U Fission}, Weiss \textit{et al.} discussed the first observation of $^{118}$Ag in 1968 \cite{Wei68}. A uranium solution was irradiated with neutrons in the Vallecitos Nuclear Test Reactor of the  U.S. Naval Radiological Defense Laboratory in San Francisco, CA. The chemically separated samples were analyzed using atomic absorption spectrometry. ``Analysis by the method of least squares gives a half-life of 5.3$^{+0.9}_{-0.7}$~s.'' This half-life is close to the accepted value of 3.76(15)~s. A previous half-life measurement of 25~s \cite{Fri67} could not be confirmed.

\subsection*{$^{119}$Ag}\vspace{-0.85cm}

$^{119}$Ag was discovered by Kawase \textit{et al.} in 1975 in their paper \textit{States in $^{119}$Cd Studied in the Decay of $^{119}$Ag} \cite{Kaw75}. $^{119}$Ag was observed at the OSIRIS mass separator in fission products from the reactor at Studsvik, Sweden. Conversion electrons, $\gamma$-rays, and $\gamma \gamma$ coincidences were recorded. ``Only one isomer of $^{119}$Ag was found in the present study, and the half-life of this, 2.1$\pm$0.1~s, is much shorter than that of its daughters products.'' An earlier measurement of a half-life of 17~s \cite{Fri67} could not be confirmed. Also, Aleklett \textit{et al.} submitted their measurement of $^{119}$Ag only two months later than Kawase \textit{et al.} \cite{Kaw75}.

\subsection*{$^{120}$Ag}\vspace{-0.85cm}

In 1971 Fogelberg \textit{et al.} reported the first observation of $^{120}$Ag in \textit{Energy Levels in $^{114,116,118,120,122}$Ca as observed in the beta decay of Ag isotopes} \cite{Fog71}. $^{120}$Ag was produced via thermal neutron fission in a uranium target at the Studsvik R2-0 reactor and separated with the OSIRIS on-line mass-separator facility. Gamma-ray singles and coincidences were measured with Ge(Li) detectors. ``The nuclides $^{120}$Ag and $^{122}$Ag have been studied for the first time...'' The measured half-life of 1.17(5)~s for the ground state agrees with the presently adapted value of 1.23(4)~s.

\subsection*{$^{121}$Ag}\vspace{-0.85cm}

Fogelberg and Hoff discovered $^{121}$Ag in 1982 as reported in \textit{Levels and Transistion Probabilities in $^{121}$Cd} \cite{Fog82}. $^{121}$Ag was produced via thermal neutron fission in an uranium target at the Studsvik R2-0 reactor and separated with the OSIRIS on-line mass-separator facility. ``Only one $\beta$-decaying state of $^{121}$Ag was found. The half-life was determined to 0.72$\pm$0.10~s which is almost an order of magnitude shorter than for any of the daughter activities.'' This half-life is included in the weighted average to determine the presently accepted value of 0.79(2)~s. It should be mentioned that Aleklett \textit{et al.} discussed $^{121}$Ag in a paper submitted four months earlier \cite{Ale82}, but since they referred to the half-life measurement of Fogelberg and Hoff as submitted we credit the latter with the discovery.

\subsection*{$^{122}$Ag}\vspace{-0.85cm}

In 1971, Fogelberg \textit{et al.} reported the first observation of $^{122}$Ag in \textit{Energy Levels in $^{114,116,118,120,122}$Ca as observed in the beta decay of Ag isotopes} \cite{Fog71}. $^{122}$Ag was produced via thermal neutron fission in an uranium target at the Studsvik R2-0 reactor and separated with the OSIRIS on-line mass-separator facility. Gamma-ray singles and coincidences were measured with Ge(Li) detectors. ``The nuclides $^{120}$Ag and $^{122}$Ag have been studied for the first time...'' Although the measured half-life is too large (1.5(5)~s) compared the currently accepted value of 520(14)~ms, we credit Fogelberg \textit{et al.} with the discovery because the coincident $\gamma$-rays of $^{122}$Cd were correctly identified.

\subsection*{$^{123}$Ag}\vspace{-0.85cm}

$^{123}$Ag was discovered by Lund and Rundstam in 1976 as reported in \textit{Delayed-neutron activities produced in fission: Mass range 122-146} \cite{Lun76}. $^{123}$Ag was produced via neutron fission in a uranium target at the Studsvik R2-0 reactor and separated with the OSIRIS on-line mass-separator facility. 30 $^3$He neutron counters were used to measure the delayed neutron activities. ``From mass formula predictions the indium and cadmium isobars of this mass are not likely to be delayed neutron precursors. Silver, on the other hand, has a positive neutron window. Consequently, it seems probable that the 0.39 sec activity is due to $^{123}$Ag.''  The half-life measurement of 390(30)~ms is close to the currently accepted value of 296(6)~ms.

\subsection*{$^{124}$Ag}\vspace{-0.85cm}

$^{124}$Ag was first correctly identified by Hill \textit{et al.} in 1984 as reported in \textit{Identification and decay of $^{124}$Ag} \cite{Hil84}. $^{124}$Ag was produced by neutron irradiation of $^{235}$U at the High Flux Beam Reactor at Brookhaven National Laboratory. The isotope was identified in the TRISTAN mass separator facility and $\gamma$-ray singles and coincidences were detected with two high-purity germanium detectors. ``We attribute the single $\gamma$ ray at 613.2 keV with a half-life of 0.17 s to come from the decay of $^{124}$Ag.'' The measured half-life of 0.17(3)~s agrees with the currently adapted value of 172(5)~ms. A previous observation of a half-life of 0.54(8)~s for $^{124}$Ag \cite{Ree83} could not be confirmed.

\subsection*{$^{125,126}$Ag}\vspace{-0.85cm}

Bernas {\it{et al.}} discovered $^{125}$Ag and $^{126}$Ag in 1994 at GSI, Germany, as reported in {\it{Projectile Fission at Relativistic Velocities: A Novel and Powerful Source of Neutron-Rich Isotopes Well Suited for In-Flight Isotopic Separation}} \cite{Ber94}. The isotopes were produced using projectile fission of $^{238}$U at 750 MeV/nucleon on a lead target. ``Forward emitted fragments from $^{80}$Zn up to $^{155}$Ce were analyzed with the Fragment Separator (FRS) and unambiguously identified by their energy-loss and time-of-flight.'' The experiment yielded 119 and 19 individual counts of $^{125}$Ag and $^{126}$Ag, respectively.

\subsection*{$^{127}$Ag}\vspace{-0.85cm}

$^{127}$Ag was first observed in 1995 by Fedoseyev \textit{et al.} and reported in \textit{Study of short-lived silver isotopes with a laser ion source} \cite{Fed95}. $^{127}$Ag was produced by proton-induced fission at the PS-Booster ISOLDE facility at CERN, Switzerland. The identification was achieved by resonance ionization using a chemically selective laser ion source. ``Decay properties of the neutron-rich isotopes $^{121-127}$Ag were studied with a neutron long-counter and a $\beta$-detector.'' The half-life was determined to be 109(25)~ms which agrees with the currently adapted half-life of 79(3)~ms.

\subsection*{$^{128}$Ag}\vspace{-0.85cm}

In 2000 Kautzsch \textit{et al.} reported the discovery of $^{128}$Ag in \textit{New states in heavy Cd isotopes and evidence for weakening of the N = 82 shell structure.} \cite{Kau00}. A pulsed 1 GeV proton beam from the CERN Proton Synchrotron Booster bombarded a thick UC${_2}$-C target and $^{128}$Ag was identified using resonance ionization laser ion sources (RILIS) at ISOLDE. The caption of Figure 1 stated ``Excerpts of ``laser-on'' $\gamma$-singles spectra for A =126 and A = 128... The 95-ms $^{126}$Ag and 58-ms $^{128}$Ag peaks are only seen in the respective early spectrum.'' This half-life corresponds to the presently accepted value of 58(5)~ms.

\subsection*{$^{129}$Ag}\vspace{-0.85cm}

$^{129}$Ag was first observed in 1995 by Fedoseyev \textit{et al.} and reported in \textit{Study of short-lived silver isotopes with a laser ion source} \cite{Fed95}. $^{129}$Ag was produced by proton-induced fission at the PS-Booster ISOLDE facility at CERN, Switzerland. The identification was achieved by resonance ionization using a chemically selective laser ion source. ``Although the LIS conditions were not optimized, we probably have already ``seen'' a $^{129}$Ag component underlying the $^{129}$In isobar; however, with too low intensity to extract a reliable half-life.''

\subsection*{$^{130}$Ag}\vspace{-0.85cm}

In 2000 Kautzsch \textit{et al.} reported the discovery of $^{130}$Ag in \textit{New states in heavy Cd isotopes and evidence for weakening of the N = 82 shell structure.} \cite{Kau00}. A pulsed 1 GeV proton beam from the CERN Proton Synchrotron Booster bombarded a thick UC${_2}$-C target and $^{128}$Ag was identified using resonance ionization laser ion sources (RILIS) at ISOLDE. ``One of them at an energy of 957 keV, which is only observed in the first time-bin and decays with an estimated half-life of about 50~ms is tentatively attributed to $^{130}$Ag decay and may represent the 2$^+ \rightarrow 0^+$ transition in neutron-magic $^{130}$Cd$_{82}$.'' This half-life is currently the only measured value for $^{130}$Ag.

\section{Summary}

The discovery of the isotopes of silver has been cataloged and the methods of their discovery discussed. The assignment of discovery was very difficult for many isotopes. The first half-life measurements for 6 silver isotopes ($^{97}$Ag, $^{100}$Ag, $^{114}$Ag, $^{118}$Ag, $^{119}$Ag, and $^{124}$Ag) were incorrect. The half-lives of $^{102-105}$Ag, and$^{115}$Ag were initially assigned to a different isotope. In addition, the half-lives of $^{104}$Ag, $^{108}$Ag, $^{110}$Ag, and $^{116}$Ag were first measured without a definite mass assignment.

\ack

This work was supported by the National Science Foundation under grants No. PHY06-06007 (NSCL) and PHY07-54541 (REU). MH was supported by NSF grant PHY05-55445. JQG acknowledges the support of the Professorial Assistantship Program of the Honors College at Michigan State University.


\newpage

\section*{EXPLANATION OF TABLE}\label{sec.eot}
\addcontentsline{toc}{section}{EXPLANATION OF TABLE}

\renewcommand{\arraystretch}{1.0}

\begin{tabular*}{0.95\textwidth}{@{}@{\extracolsep{\fill}}lp{5.5in}@{}}
\textbf{TABLE I.}
	& \textbf{Discovery of Silver Isotopes }\\
\\

Isotope & Silver isotope \\
Author & First author of refereed publication \\
Journal & Journal of publication \\
Ref. & Reference \\
Method & Production method used in the discovery: \\
 & FE: fusion evaporation \\
 & LP: light-particle reactions (including neutrons) \\
 & MS: mass spectroscopy \\
 & PN: photonuclear reactions \\
 & NC: neutron-capture reactions \\
 & SP: spallation \\
 & NF: neutron-induced fission \\
 & CPF: charged-particle induced fission \\
 & PF: projectile fragmentation or projectile fission \\
Laboratory & Laboratory where the experiment was performed\\
Country & Country of laboratory\\
Year & Year of discovery \\
\end{tabular*}
\label{tableI}

\newpage
\datatables

\setlength{\LTleft}{0pt}
\setlength{\LTright}{0pt}


\setlength{\tabcolsep}{0.5\tabcolsep}

\renewcommand{\arraystretch}{1.0}


\begin{longtable}[c]{%
@{}@{\extracolsep{\fill}}r@{\hspace{5\tabcolsep}} llllllll@{}}
\caption[Discovery of Silver Isotopes]%
{Discovery of Silver isotopes}\\[0pt]
\caption*{\small{See page \pageref{tableI} for Explanation of Tables}}\\
\hline
\\[100pt]
\multicolumn{8}{c}{\textit{This space intentionally left blank}}\\
\endfirsthead
Isotope & Author & Journal & Ref. & Method & Laboratory & Country & Year \\
$^{93}$Ag & M. Hencheck & Phys. Rev. C & Hen94 & PF & Michigan State & USA &1994 \\
$^{94}$Ag & M. Hencheck & Phys. Rev. C & Hen94 & PF & Michigan State & USA &1994 \\
$^{95}$Ag & M. Hencheck & Phys. Rev. C & Hen94 & PF & Michigan State & USA &1994 \\
$^{96}$Ag & W. Kurcewicz & Z. Phys. A & Kur82 & FE & Darmstadt & Germany &1982 \\
$^{97}$Ag & M. Huyse & Z. Phys. A & Huy78 & FE & Louvain-la-Neuve & Belgium &1978 \\
$^{98}$Ag & M. Huyse & Z. Phys. A & Huy78 & FE & Louvain-la-Neuve & Belgium &1978 \\
$^{99}$Ag & H. Bakhru & Nucl. Phys. A & Bak67 & FE & Yale & USA &1967 \\
$^{100}$Ag & D.J. Hnatowich & J. Inorg. Nucl. Chem. & Hna70 & SP & CERN & Switzerland &1970 \\
$^{101}$Ag & F.D.S. Butement & J. Inorg. Nucl. Chem. & But66 & SP & Liverpool & UK &1966 \\
$^{102}$Ag & O. Ames & Phys. Rev. & Ame60 & LP & Princeton & USA &1960 \\
$^{103}$Ag & B.C. Haldar & Phys. Rev. & Hal54 & LP & Rochester & USA &1954 \\
$^{104}$Ag & F.A. Johnson & Can. J. Phys. & Joh55 & LP & McGill & Canada &1955 \\
$^{105}$Ag & T. Enns & Phys. Rev. & Enn39 & LP & Rochester & USA &1939 \\
$^{106}$Ag & W. Bothe & Naturwiss. & Bot37 & PN & Heidelberg & Germany &1937 \\
$^{107}$Ag & F.W. Aston & Phil. Mag. & Ast24 & MS & Cambridge & UK &1924 \\
$^{108}$Ag & W. Bothe & Naturwiss. & Bot37 & PN & Heidelberg & Germany &1937 \\
$^{109}$Ag & F.W. Aston & Phil. Mag. & Ast24 & MS & Cambridge & UK &1924 \\
$^{110}$Ag & W. Bothe & Naturwiss. & Bot37 & NC & Heidelberg & Germany &1937 \\
$^{111}$Ag & J.D. Kraus & Phys. Rev. & Kra37 & LP & Michigan & USA &1937 \\
$^{112}$Ag & M.L. Pool & Phys. Rev. & Poo38 & LP & Michigan & USA &1938 \\
$^{113}$Ag & R.B. Duffield & Phys. Rev. & Duf49 & PN & Illinois & USA &1949 \\
$^{114}$Ag & J. M. Alexander & Phys. Rev. & Ale58 & CPF & MIT & USA &1958 \\
$^{115}$Ag & R.B. Duffield & Phys. Rev. & Duf49 & PN & Illinois & USA &1949 \\
$^{116}$Ag & J. M. Alexander & Phys. Rev. & Ale58 & CPF & MIT & USA &1958 \\
$^{117}$Ag & J. M. Alexander & Phys. Rev. & Ale58 & CPF & MIT & USA &1958 \\
$^{118}$Ag & H.V. Weiss & Phys. Rev. & Wei68 & NF & U.S. Naval Rad. Def. Lab. & USA &1967 \\
$^{119}$Ag & Y. Kawase & Nucl. Phys. A & Kaw75 & NF & Studsvik & Sweden &1975 \\
$^{120}$Ag & B. Fogelberg & Phys. Lett. B & Fog71 & NF & Studsvik & Sweden &1971 \\
$^{121}$Ag & B. Fogelberg & Nucl. Phys. A & Fog82 & NF & Studsvik & Sweden &1982 \\
$^{122}$Ag & B. Fogelberg & Phys. Lett. B & Fog71 & NF & Studsvik & Sweden &1971 \\
$^{123}$Ag & E. Lund & Phys. Rev. C & Lun76 & NF & Studsvik & Sweden &1976 \\
$^{124}$Ag & J.C. Hill & Phys. Rev. C & Hil84 & NF & Brookhaven & USA &1984 \\
$^{125}$Ag & M. Bernas & Phys. Lett. B & Ber94 & PF & Darmstadt & Germany &1994 \\
$^{126}$Ag & M. Bernas & Phys. Lett. B & Ber94 & PF & Darmstadt & Germany &1994 \\
$^{127}$Ag & V.N. Fedoseyev & Z. Phys. A & Fed95 & SP & CERN & Switzerland &1995 \\
$^{128}$Ag & T. Kautzsch & Eur. Phys. J. A & Kau00 & SP & CERN & Switzerland &2000 \\
$^{129}$Ag & V.N. Fedoseyev & Z. Phys. A & Fed95 & SP & CERN & Switzerland &1995 \\
$^{130}$Ag & T. Kautzsch & Eur. Phys. J. A & Kau00 & SP & CERN & Switzerland &2000 \\
\end{longtable}

\newpage


\normalsize

\begin{theDTbibliography}{1956He83}
\bibitem[Ale58]{Ale58t} J.M. Alexander, U. Schindewolf, and C.D. Corvell, Phys. Rev. {\bf 111}, 228 (1958)
\bibitem[Ame60]{Ame60t} O. Ames, A.M. Bernstein, M.H. Brennan, R.A. Haberstroh, and D.R. Hamilton, Phys. Rev. {\bf 118}, 1599 (1960)
\bibitem[Ast24]{Ast24t} F.W. Aston, Phil. Mag. {\bf 47}, 385 (1924)
\bibitem[Bak67]{Bak67t} H. Bakhru, R.I. Morse, and I.L. Preiss, Nucl. Phys. A {\bf 100}, 145 (1967)
\bibitem[Ber94]{Ber94t} M. Bernas, S. Czajkowski, P. Armbruster, H. Geissel, Ph. Dessagne, C. Donzaud, H.-R. Faust, E. Hanelt, A. Heinz, M. Hesse, C. Kozhuharov, Ch. Miehe, G. M\"unzenberg, M. Pf\"utzner, C. R\"ohl, K.-H. Schmidt, W. Schwab, C. St\'ephan, K. S\"ummerer, L. Tassan-Got, and B. Voss, Phys. Lett. B {\bf 331}, 19 (1994)
\bibitem[Bot37]{Bot37t} W. Bothe and W. Gentner, Naturwiss. {\bf 25}, 126 (1937)
\bibitem[But66]{But66t} F.D.S. Butement and M.Y. Mirza, J. Inorg. Nucl. Chem. {\bf 28}, 303 (1966)
\bibitem[Duf49]{Duf49t} R.B. Duffield and J.D. Knight, Phys. Rev. {\bf 75}, 1613 (1949)
\bibitem[Enn39]{Enn39t} T. Enns, Phys. Rev. {\bf 56}, 872 (1939)
\bibitem[Fed95]{Fed95t} V.N. Fedoseyev, Y. Jading, O.C. Jonsson, R. Kirchner, K.-L. Kratz, M. Krieg, E. Kugler, J. Lettry, T. Mehren, V.I. Mishin, H.L. Ravn, T. Rauscher, H.L. Ravn, F. Scheerer, O. Tengblad, P. Van Duppen, A. Wohr, and the ISOLDE Collaboration, Z. Phys. A {\bf 353}, 9 (1995)
\bibitem[Fog71]{Fog71t} B. Fogelberg, A. Backlin, and T. Nagarajan, Phys. Lett. B {\bf 36}, 334 (1971)
\bibitem[Fog82]{Fog82t} B. Fogelberg and P. Hoff, Nucl. Phys. A {\bf 391}, 445 (1982)
\bibitem[Hal54]{Hal54t} B.C. Haldar and E.O. Wiig, Phys. Rev. {\bf 94}, 1713 (1954)
\bibitem[Hen94]{Hen94t} M. Hencheck, R.N. Boyd, M. Hellstr\"om, D.J. Morrissey, M.J. Balbes, F.R. Choupek, M. Fauerbach, C.A. Mitchell, R. Pfaff, C.F. Powell, G. Raimann, B.M. Sherrill, M. Steiner, J. Vandegriff, and S.J. Yennello, Phys. Rev. C {\bf 50}, 2219 (1994)
\bibitem[Hil84]{Hil84t} J.C Hill, F.K. Wohn, Z. Berant, R.L. Gill, R.E. Chrien, C. Chung, and A. Aprahamian, Phys. Rev. C {\bf 29}, 1078 (1984)
\bibitem[Hna70]{Hna70t} D.J. Hnatowich, E. Hagebo, A. Kjelberg, R. Mohr, and P. Patzelt, J. Inorg. Nucl. Chem. {\bf 32}, 3137 (1970)
\bibitem[Huy78]{Huy78t} M. Huyse, K. Cornelis, G. Dumont, G. Lhersonneau, J. Verplancke, and W.B. Walters, Z. Phys. A {\bf 288}, 107 (1997)
\bibitem[Joh55]{Joh55t} F.A. Johnson, Can. J. Phys. {\bf 33}, 841 (1955)
\bibitem[Kau00]{Kau00t} T. Kautzsch, W.B. Walters, M. Hannawald, K.-L. Kratz, V.I. Mishin, V.N. Fedoseyev, W. B\"ohmer, Y. Jading, P. Van Duppen, B. Pfeiffer, A. W\"ohr, P. M\"oller, I. Kl\"ockl, V. Sebastian, U. K\"oster, M. Koizumi, J. Lettry, H.L. Ravn, and the ISOLDE Collaboration, Eur. Phys. J. A {\bf 9}, 201 (2000)
\bibitem[Kaw75]{Kaw75t} Y. Kawase, B. Fogelberg, J. McDonald, and A. Backlin, Nucl. Phys. A {\bf 241}, 237 (1975)
\bibitem[Kra37]{Kra37t} J.D. Kraus and J.M. Cork, Phys. Rev. {\bf 52}, 763 (1937)
\bibitem[Kur82]{Kur82t} W. Kurcewicz, E.F. Zganjar, R. Kirchner, O. Klepper, E. Roeckl, P. Komninos, E. Nolte, D. Schardt, and P. Tidemand-Petersson, Z. Phys. A {\bf 308}, 21 (1982)
\bibitem[Lun76]{Lun76t} E. Lund and G. Runstam, Phys. Rev. C {\bf 13}, 1544 (1976)
\bibitem[Poo38]{Poo38t} M.L. Pool, Phys. Rev. {\bf 53}, 116 (1938)
\bibitem[Wei68]{Wei68t} H.V. Weiss, J.M. Fresco, and W.L. Reichert, Phys. Rev. {\bf 172}, 1266 (1968)

\end{theDTbibliography}

\end{document}